# Pb-Radioactivity in superheavy elements

Sushil Kumar*
Physics Department, Chitkara University, Solan-174103, INDIA
* email: sushilk17@gmail.com

## Introduction

A study on exotic natural radioactivity of emitting nuclei heavier than α-particle, known as cluster radioactivity. Theoretically, it was predicted in 1980 by Săndulescu, Poenaru and Greiner [1]. Later it was confirmed experimentally by Rose and Jones in 1984 [2] via the $^{14}$C decay from $^{223}$Ra nucleus. Since then many other $^{14}$C-decays from other radioactive nuclei ($^{221}$Fr, $^{221,222,224,226}$Ra, $^{223,225}$Ac) and some other neutron rich clusters, such as $^{20}$O, $^{23}$F, $^{22,24-26}$Ne, $^{28,30}$Mg and $^{32,34}$Si, have been observed experimentally for the ground state decays of translead $^{226}$Th to $^{242}$Cm parents [3, 4], which all decay with the doubly closed shell daughter $^{208}$Pb (Z=82, N=126) or its neighboring nuclei. Recently, Poenaru et. al., studied the heavy particle radioactivity (Ze >28) of superheavy elements using the Analytical Super Asymmetric Fission (ASAF) model [5].

During the last few years, advancement in the technology has opened the door for many new superheavy elements. Recently, the isotopes $^{293}$117 and $^{294}$117 were produced in fusion reactions between $^{48}$Ca and $^{249}$Bk nuclei [6]. These superheavy elements mainly undergo sequential alpha-decays, which ends with spontaneous fission (SF). The cluster decay study of superheavy nuclei is a subject of interest and a possibility looks with lead as a daughter or its neighbor in this region after the small half-lives and cross sections of these nuclei [6].

## Preformed Cluster Model

The preformed cluster model (PCM) [7] uses the dynamical collective coordinates of mass and charge asymmetries $\eta$ and $\eta_z$ on the basis of Quantum Mechanical Fragmentation Theory. The decay constant $\lambda$ in PCM is defined as

$$\lambda = \frac{\ln 2}{T_{1/2}} = P_0 \nu_0 P \qquad (1)$$

Here $P_0$ is the cluster preformation probability and P is the barrier penetrability which refer, respectively, to the $\eta$- and R- motions. $\nu_0$ is the barrier assault frequency. $P_0$ are the solutions of the stationary Schrodinger equation in $\eta$,

$$\left\{-\frac{\hbar^2}{2\sqrt{B_{\eta\eta}}}\frac{\partial}{\partial\eta}\frac{1}{\sqrt{B_{\eta\eta}}}\frac{\partial}{\partial\eta}+V_R(\eta)\right\}\psi^{(\nu)}(\eta)=E^{(\nu)}\psi^{(\nu)}(\eta) \qquad (2)$$

Which on proper normalization are given as

$$P_0 = \sqrt{B_{\eta\eta}}\left|\psi^{(0)}(\eta(A_i))\right|^2 \left(\frac{2}{A}\right) \qquad (3)$$

The fragmentation potential ($V_R(\eta)$ in eq (2) is calculated simply as the sum of Coulomb interaction, the nuclear proximity potential and the ground state binding energies of two nuclei:

$$V(R_a,\eta) = -\sum_{i=1}^{2}B(A_i,Z_i)+\frac{Z_1Z_2e^2}{R_a}+V_P \qquad (4)$$

With B's taken from the 2003 experimental compilation of Audi et al and from the 1995 calculations of Moller et al. Thus, full shell effects are contained in our calculations that come from the experimental and/or calculated binding energies. The WKB tunneling probability calculated is P= $P_iP_b$ with

$$P_i = \exp\left[-\frac{2}{\hbar}\int_{R_a}^{R_i}\{2\mu[V(R)-V(R_i)]\}^{1/2}dR\right] \qquad (5)$$

$$P_b = \exp\left[-\frac{2}{\hbar}\int_{R_i}^{R_b}\{2\mu[V(R)-Q]\}^{1/2}dR\right] \qquad (6)$$

These integrals are solved analytically for $R_b$, the second turning point, defined by $V(R_b)$=Q-value for the ground- state decay.

The assault frequency $\nu_0$ is given siply as

$$\nu_0 = \left(2E_2/\mu\right)^{1/2}/R_0 \qquad (7)$$

With $E_2=(A_1/A)Q$, the kinetic energy of lighter fragment, for the Q- value shared between the two products as inverse of their masses.

## Calculation and Results

In this study, previous cluster decay calculation [8] within PCM model considered with atomic number ($2 < Z_{cluster} < 20$) now is extended up to the heavier cluster ($Z_c^{max} = Z_p - 82$), to get information about the most probable cluster and a doubly magic daughter around $^{208}$Pb for the island of superheavy elements. The PCM model [7] is used for these decay calculations. The decay characteristic of $^{288}$114 parent nucleus shows that $^{80}$Ge is most probable cluster with $^{208}$Pb daughter, shown in Fig. 1. The result of PCM model calculation is same as the results of calculation in [5].

The results, first time for the calculation of Pb-Radioactivity looks favorably for the cluster decay studies in superheavy elements as in the actinide region. The results are shown in Table 1 with the alpha decay half-lives of various superheavy parents.

**Table 1:** The decay half-lives of various super heavy elements.

| Parent | Daughter | Cluster | Half-life $Log_{10}T_{1/2}$ |
|---|---|---|---|
| $^{288}$114 | $^{284}$112 | $^4$He | 3.196 |
| | $^{210}$Pb | $^{78}$Ge | 1.972 |
| | $^{208}$Pb | $^{80}$Ge | -2.955 |
| | $^{207}$Pb | $^{81}$Ge | -1.178 |
| | $^{206}$Pb | $^{82}$Ge | -1.713 |
| $^{288}$115 | $^{284}$113 | $^4$He | 0.868 |
| | $^{206}$Pb | $^{82}$As | -5.442 |
| $^{287}$115 | $^{284}$113 | $^4$He | 3.003 |
| | $^{204}$Pb | $^{83}$As | -2.206 |
| $^{294}$117 | $^{290}$115 | $^4$He | 3.891 |
| | $^{208}$Pb | $^{86}$Br | -5.45 |
| | $^{207}$Pb | $^{87}$Br | -4.787 |
| $^{293}$117 | $^{289}$115 | $^4$He | 0.009 |
| | $^{206}$Pb | $^{87}$Br | -8.275 |

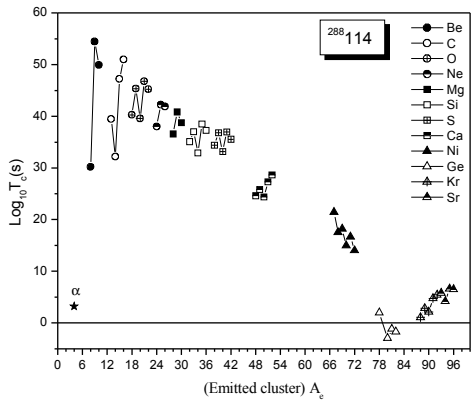

**Fig. 1** $\log_{10}T_{1/2}$ (s) versus emitted cluster mass number ($A_2$) for the $^{288}$114 parent nucleus.